\documentclass[aps,prb,twocolumn,showpacs,floatfix,superscriptaddress]{revtex4-1}

\usepackage{gensymb}
\usepackage{graphicx}
\usepackage{xcolor}
\usepackage{float}
\usepackage[utf8x]{inputenc}
\usepackage{amsmath}
\usepackage{subfig}
\usepackage{hyperref}
\usepackage{lipsum}

\begin{document}

\title
{Comparison of Solar Cell Efficiencies of Black Phosphorus and Silicon at the Nano and Micro Scales from First-Principles Calculations}  
\author{Burak Ozdemir}
\email{burkzdemir@gmail.com}
\affiliation{Department of Physics, Faculty of Science, University of Ostrava, 30. dubna 22, 70103 Ostrava, Czech Republic}
\date\today

\begin{abstract}
Density functional theory and many-body (GW+BSE) calculations of transmittance, absorbance, and reflectance are performed on silicon and black phosphorus (BP). We find that a damping value of 0.01 used in the dielectric function calculation is the optimal for calculating the solar cell efficiency of Si. Our calculations indicate that the solar cell efficiency of a 100 $\mu$m thick Si slab is 27.4\% while the efficiency of BP for the same thickness is 2.33\% and 1.94\% for light polarized along the zigzag and armchair directions, respectively. For 100 nm thick materials, we obtain that Si presents a 0.8\% solar cell efficiency and BP exhibits a 0.14\% and 1.02\% efficiency for light polarized along the the zigzag and armchair directions, respectively, indicating that BP performs better than Si at these small scales. Our results underscore the important effect of the material thickness on solar cell efficiencies. 
\end{abstract}

\maketitle

\section{Introduction}
Due to its importance in solar cell applications, Si has been studied very well experimentally. The highest experimental solar cell efficiency of Si reported in the literature is 24.7\% for a 98 $\mu$m thick crystal.\cite{taguchi201324} Experimental values of the transmittance and reflectance of Si have also been reported in the literature,\cite{green2008self,serpenguzelsilicon} including the transmittance of Si below the band gap at 55\%.\cite{serpenguzelsilicon} However, to date, there is no density functional theory (DFT) study of the optical properties of Si such as absorbance, reflectance, and transmittance. 

Black phosphorus (BP), on the other hand, presents a band gap of only 0.33 eV and, therefore, considering the Shockley-Quiesser limit, we expect a solar cell efficiency of only 5\% for BP. Although BP has low efficiency for single-junction solar cells, it could be interesting to explore its potential as a part of a tandem solar cell with improved overall efficiency. While the dielectric function of BP has been reported in the literature\cite{tran2014layer}, its thickness-dependent absorbance and reflectance have not been studied.

In this work, we perform density functional theory and G$_0$W$_0$+BSE calculations in order to study the optical properties and the thickness dependent solar cell efficiency of Si and BP.

\section{Methods}
Self-consistent DFT calculations using a plane-wave basis set are carried out by using the Quantum-Espresso software.\cite{giannozzi2009quantum} We use the semi-local density approximation (PBE) for the exchange-correlation interaction. Norm-conserving and ultrasoft pseudopotentials are used to replace core electrons for Si and BP, respectively. The kinetic energy cut-off for the wavefunctions is 50 Ry for Si and 60 Ry for BP and the charge density cutoff is 600 Ry for BP. Total energy and structure optimizations with stress minimization calculations are carried out with a Monkhorst-Pack grid \cite{monkhorst1976special} of 8$\times$6$\times$3 and 4$\times$4$\times$4 and density of states calculations are carried out with 16$\times$12$\times$6 and 16$\times$16$\times$16 grid for BP and Si, respectively. Total energy, stress, and force convergence thresholds of 10$^{-8}$ Ry, 0.5 kbar, and 10$^{-3}$ Ry/Bohr are used, respectively, for the structural optimization using the Broyden-Flether-Goldfarb-Shanno algorithm. The real and imaginary parts of the dielectric function are calculated with the Yambo code.\cite{marini2009yambo} Since these systems are semiconductors, we use the G$_0$W$_0$+BSE method to calculate the dielectric function. The Bethe-Salpeter equation (BSE) can be reduced to an eigenvalue problem,\cite{marini2009yambo}
\begin{align}
H_{\underset{nn'\vec k}{mm'\vec k'}}=&(\epsilon_{n\vec k}-\epsilon_{n'\vec k})\delta_{nm}\delta_{n'm'}\delta_{\vec k\vec k'}\\
&+(f_{n'\vec k}-f_{n\vec k})[2\bar{V}_{\underset{nn'\vec k}{mm'\vec k'}}-W_{\underset{nn'\vec k}{mm'\vec k'}}],
\end{align}
where $W$ is the electron-electron scattering term, $\bar{V}$ is the exchange interaction, $f$ is the occupation factor, $\epsilon$ is the Kohn-Sham energy.
The macroscopic dielectric function reads;
\begin{align}
\epsilon_M(\omega)=&1-\lim_{\vec q\rightarrow 0}\frac{8\pi}{|\vec q|^2\Omega N_q}\sum_{nn'\vec k}\sum_{mm'\vec k'}\rho^*_{n'n\vec k}(\vec q,\vec G)\rho_{m'm\vec k'}(\vec q,\vec G')\\
&\times\sum_\lambda\frac{A_{n'n\vec k}^\lambda(A_{m'm\vec k'}^\lambda)^*}{\omega-E_\lambda},
\label{egw}
\end{align}
where $A_{n'n\vec k}^\lambda=\langle n'n\vec k|\lambda\rangle$ are the eigenvectors of $H$, $\rho_{m'm\vec k'}(\vec q,\vec G')=\langle m'\vec k'|e^{i(\vec q +\vec G')}|m\vec k'-\vec q\rangle$, and $\vec G$ are the reciprocal lattice vectors. 
Transmission (t) and reflection (r) coefficients for the systems are obtained from the dielectric function according to Fresnel's equations for thin films;

\begin{equation}
    r=\frac{r_{01}+r_{01}e^{-2i\beta}}{1-r_{01}r_{12}e^{-2i\beta}},\quad t=\frac{t_{01}t_{12}e^{-i\beta}}{1-r_{01}r_{12}e^{-2i\beta}},
\label{efres}
\end{equation}
where $r_{01}=\frac{n_0-n_1}{n_1+n_0}$, $n_0$ and $n_1$ are the refractive indices,  $t_{01}=1+r_{01}$\cite{pedrotti}, $\beta$ is the phase difference due to the optical path length and it also involves the attenuation coefficient ($\beta=\omega n_1 d/c+i\omega \epsilon_2 d/(2n_1c)$\cite{dressel}, where $d$ is the thickness of the material). Transmittance and reflectance are obtained by taking the absolute square of the transmission and reflection coefficients; $T=|t|^2$ and $R=|r|^2$.  By using this method, we have successfully calculated transmission and reflectance of few-layer graphene in good agreement with experiments.\cite{wan2017tunable} Finally, the absorbance of the slab is calculated as $A=1-T-R$. 

Solar cell efficiencies are calculated according to the equations\cite{ozdemir2020thickness};

\begin{align}
\eta&=\frac{FFV_{oc}J_{sc}}{P_0} \\
\eta&=\frac{FFV_{oc}\int_{0.31}^{4.10}A\frac{J_{ph}(\hbar w)}{\hbar w}d(\hbar w)}{\int_{0.31}^{4.10}J_{ph}(\hbar w)d(\hbar w)},
\label{esol}
\end{align}
where FF is the filling factor, $V_{oc}$ is the open-circuit voltage, $A$ is the absorbance, J$_{ph}(\hbar w)$ is the AM1.5 solar energy flux (Wm$^{-2}$eV$^{-1}$) at the photon energy $\hbar w$. Integrals are calculated between the limits of 0.31 eV and 4.1 eV 
as the
AM1.5 solar energy flux data lie in between these values with a total of 1000 $W/m^2$ of irradiance. The AM1.5 data is taken from the National Renewable Energy Laboratory's (NREL) website (https://www.nrel.gov/grid/solar-resource/spectra-am1.5.html). In the literature, 100\% absorption above the band gap of the material is assumed, however, here, we consider the absorption of the material by multiplying J$_{ph}$ with absorbance ($A$) since not all the light is absorbed above the band gap due to reflection and the variation of absorbance due to different excitation probabilities.
Moreover, the lower limit of the short-circuit integral is taken to be 0.31 eV instead of the band gap of the material since absorption below the band gap is possible when considering a finite temperature. 
Therefore, with these considerations, we can update the theory of Shockley-Quiesser limit. FF can then be calculated for an ideal solar cell from the equation\cite{de1983fill};
\begin{equation}
    FF(v_{oc})=\biggl(1-\frac{ln( v_{oc})}{v_{oc}}\biggr)\biggl(1-\frac{1}{v_{oc}}\biggr)(1-e^{-v_{oc}})^{-1},
\end{equation}
where $v_{oc}=qV_{oc}/kT$. $V_{oc}$ is defined as\cite{de1983fill};
\begin{equation}
    V_{oc}=\frac{nkT}{q}ln\biggl(\frac{J_L}{J_0}+1\biggr)
\end{equation}
where, $T$ is the solar cell temperature taken as 300 K, $n$ is the non-ideality factor, $J_L=J_{sc}$ is the light-generated current, and $J_0$ is the dark saturation current which can be calculated with the equation\cite{baruch1995some}

\begin{equation}
    J_0=\frac{q}{k}\frac{15\sigma}{\pi^4}T^3\int_u^\infty\frac{x^2}{e^x-1}dx
\end{equation}
where   $q$   is   the   electronic   charge, $\sigma$ is   the   Stefan–Boltzmann constant, $k$ is Boltzmann constant, $T$ is the temperature and $u=E_g/kT$ ($E_g$ is the band gap).

\section{Results \& Discussion}
\begin{figure*}[t]
\centering
\subfloat[]{\label{feps5}\includegraphics[angle=0, scale=0.2]{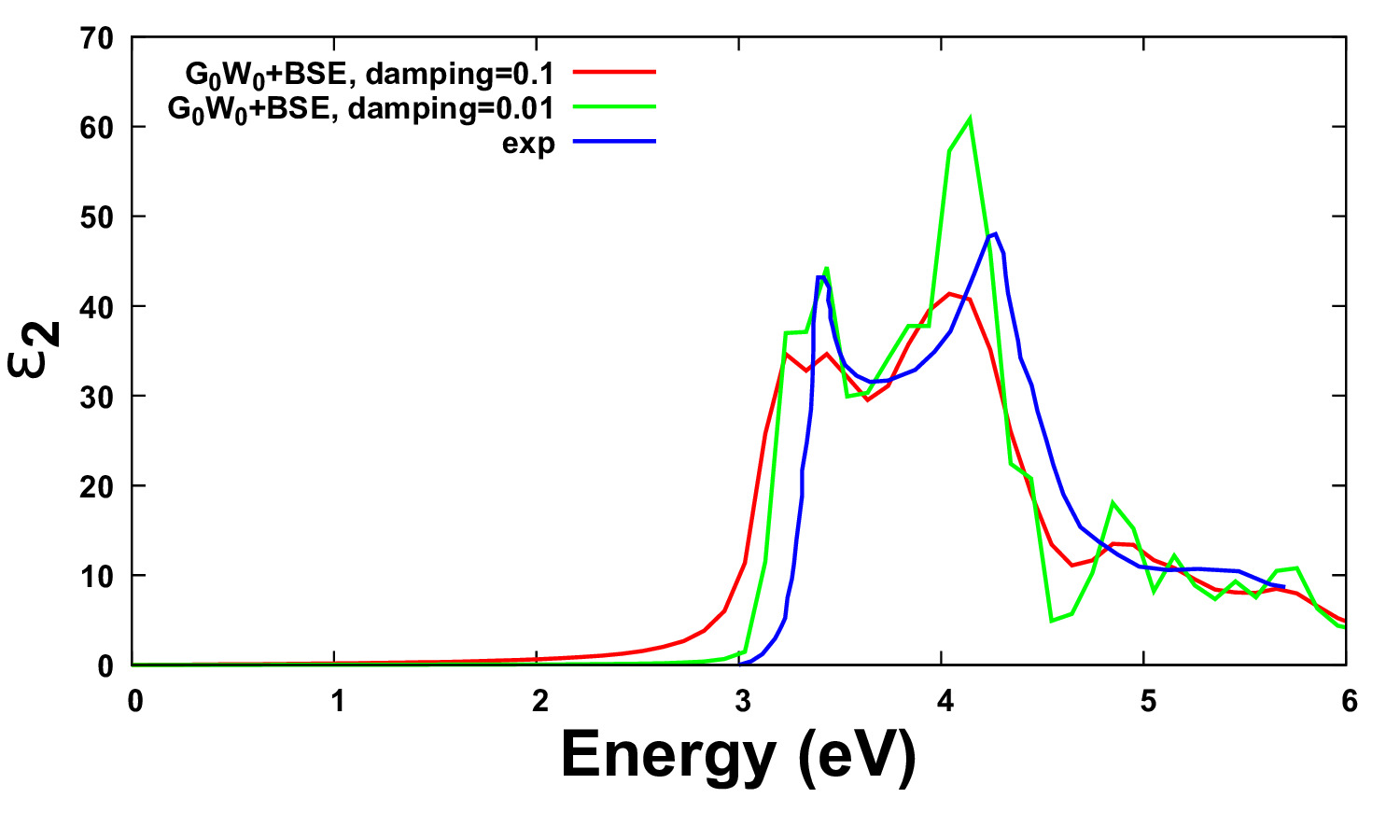}}
\caption{Comparison of converged $\epsilon_2$ of Si from a G$_0$W$_0$+BSE calculation (using two different damping values) and experiment.\cite{sottile2003macroscopic} }
\label{feps}
\end{figure*}

A convergence study of the dielectric function of Si is carried out. We determine that 2 valence bands and 3 conduction bands are enough for convergence; a k-point grid of 26$\times$26$\times$26 points is used. The reduced block size (energy cut-off of the response function) is converged at 250 and a total number of bands is converged at 200. Final dielectric functions with two different damping values are shown in Fig. \ref{feps} and compared to the experiment.

The experimental band gap of Si at 300 K is 1.12~eV. 
We obtain a PBE band gap of 0.61 eV which is improved to 1.21 eV with G$_0$W$_0$.\cite{xia2014rediscovering,precker2002experimental} These results are comparable to previous calculations.\cite{PhysRevB.87.245401,PhysRevLett.89.126401}

The experimental band gap of BP at 300 K is 0.33 eV\cite{PhysRev.92.580}, whereas the PBE gap value is 0.17 eV according to our calculations. The band gap further increases to 0.74 eV with the G$_0$W$_0$ method\cite{castellanos2015black}. 

Here, we explore thickness-dependent solar cell efficiency of Si and BP at the nano and micrometer scales. This efficiency depends, among other things, on the band gap of the material. The lowest thickness we explore is 1 nm and we assume that a nm thick silicon crystal presents the same band gap as bulk silicon. As reported previously, for black phosphorus, the band gap quickly converges to its bulk value after a few phosphorene layers are stacked on top of each other. 
\cite{tran2014layer} 

\begin{figure*}[t]
\centering
\subfloat[]{\label{fopt1}\includegraphics[angle=0, scale=0.15]{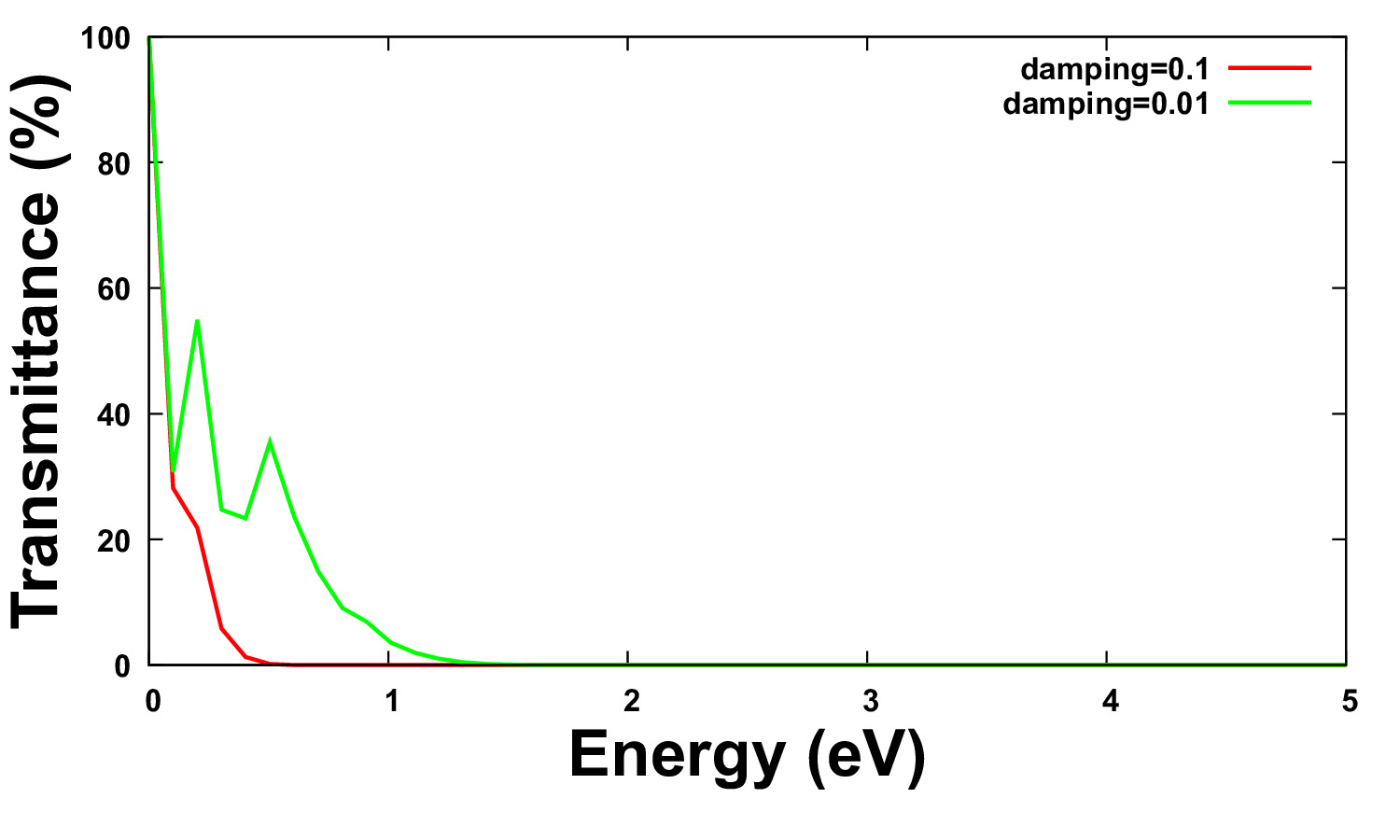}}
\subfloat[]{\label{fopt4}\includegraphics[angle=0, scale=0.15]{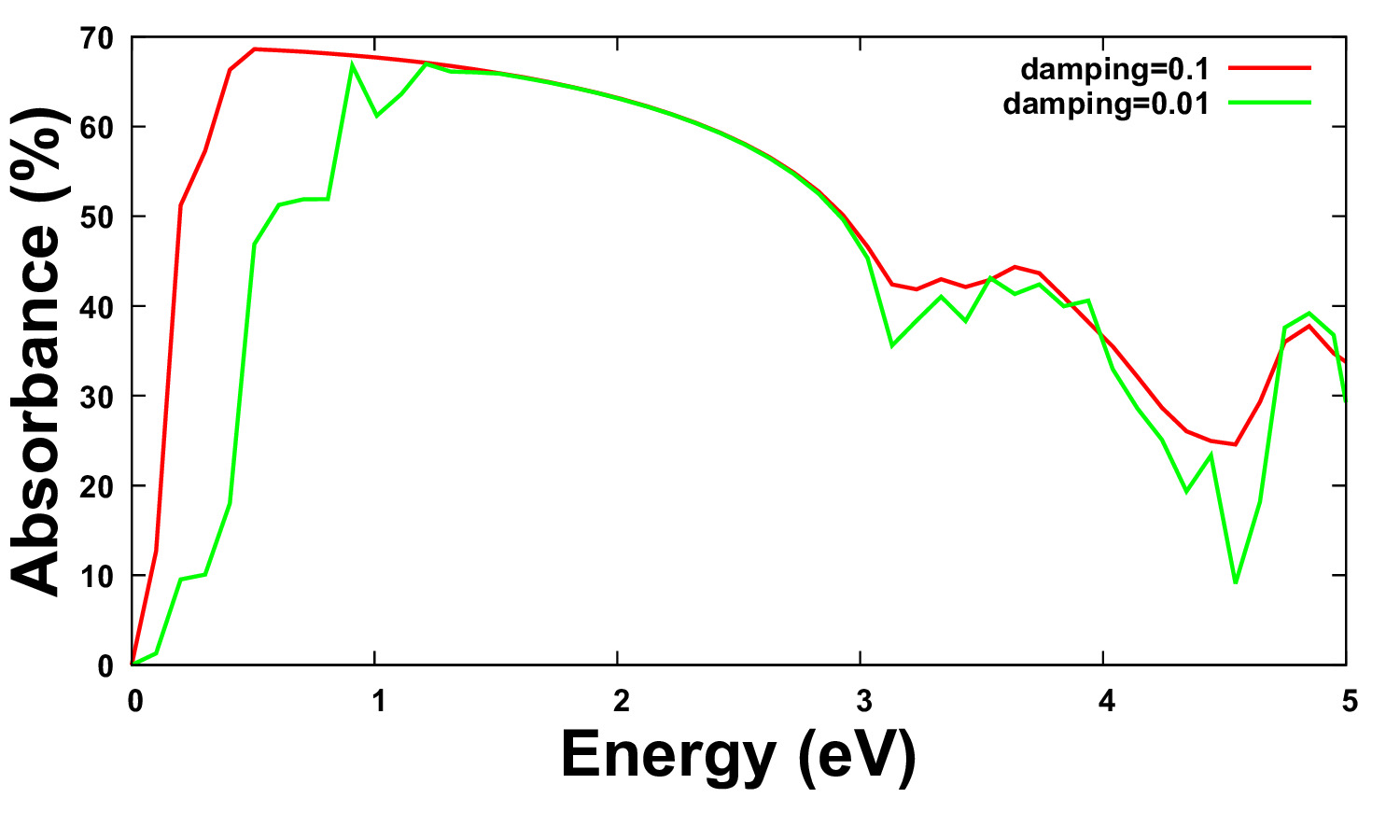}}\\
\subfloat[]{\label{fopt3}\includegraphics[angle=0, scale=0.15]{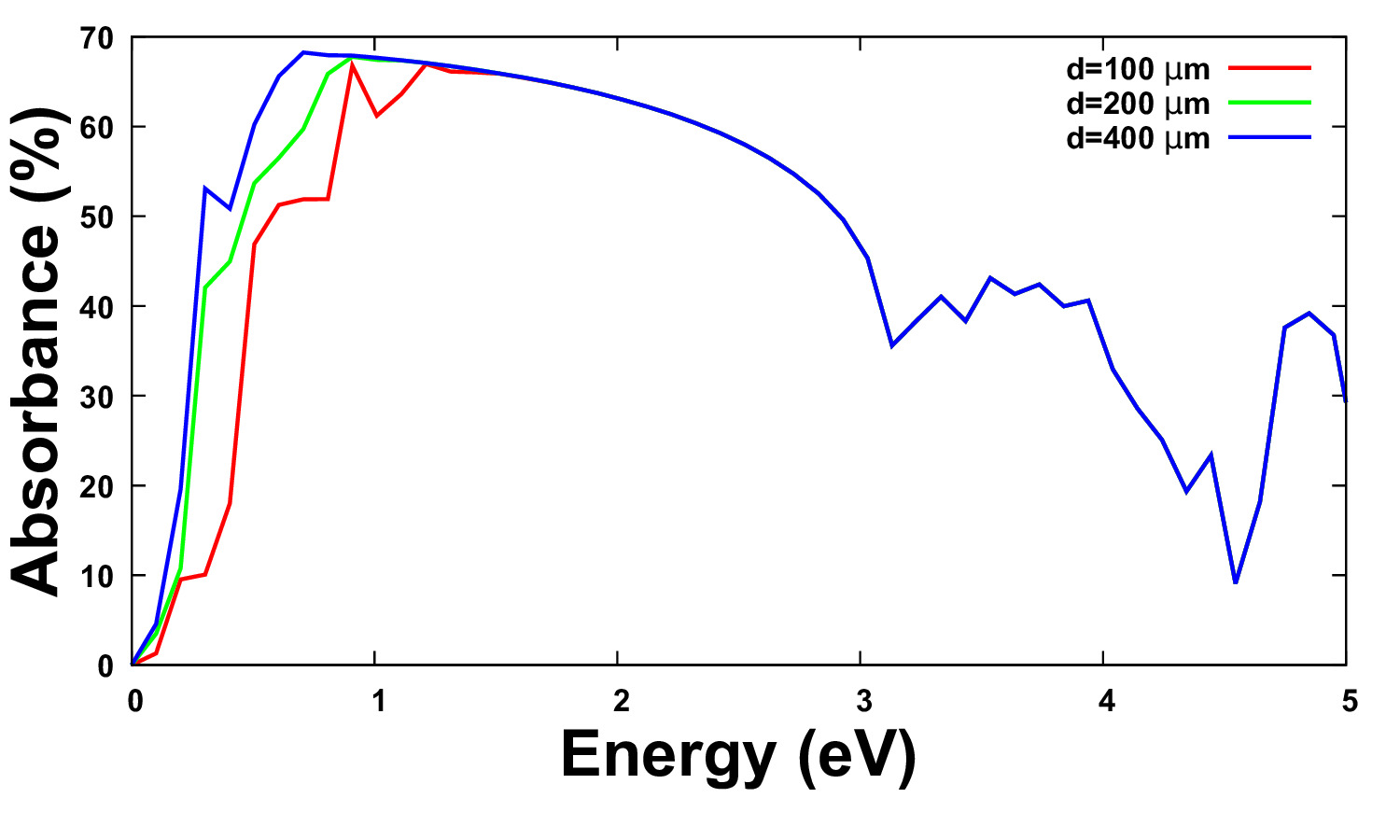}}
\subfloat[]{\label{fopt2}\includegraphics[angle=0, scale=0.15]{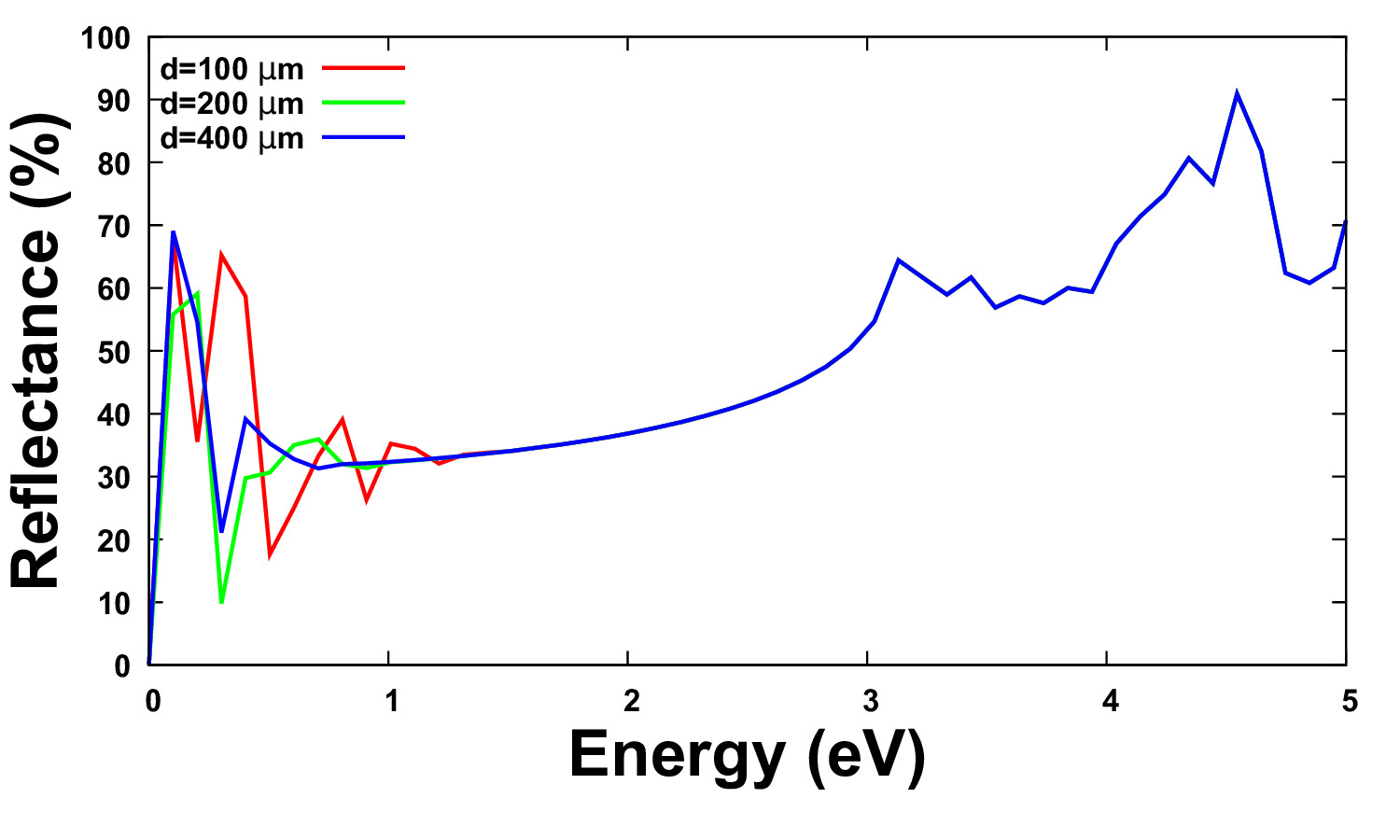}}\\
\subfloat[]{\label{fopt5}\includegraphics[angle=0, scale=0.15]{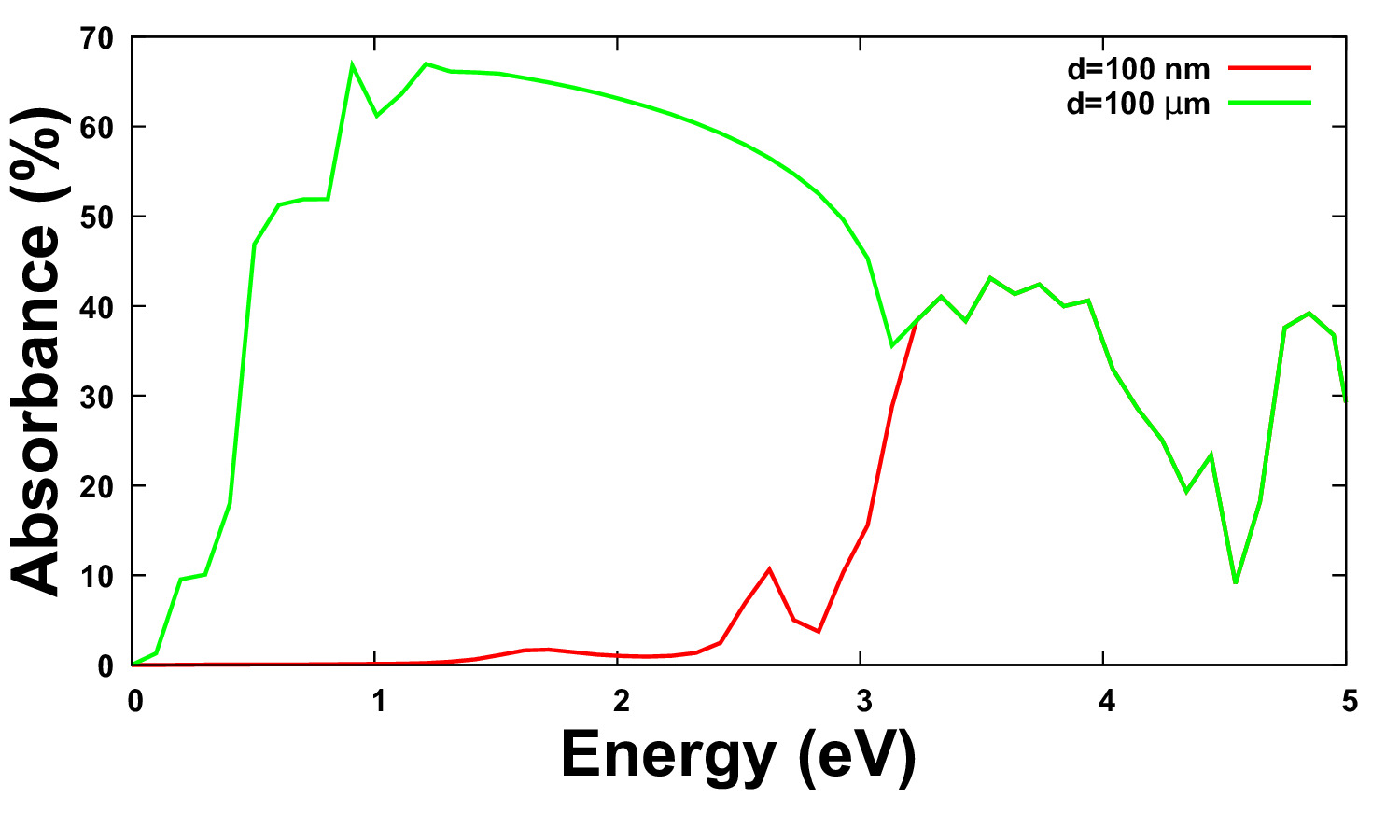}}
\caption{(a) Transmittance of 100 $\mu$m thick Si for two different damping values, (b) absorbance for 100 $\mu$m thick Si for two different damping values, (c) absorbance and (d) reflectance for three different material thicknesses at the micrometer scale, and (e) absorbance comparison of 100 nm and 100 $\mu$m thicknesses. Damping=0.01 is used in (c), (d), (e), and (f).}
\label{fopt}
\end{figure*}

The experimental reflectance of Si lies between 30-40\% at 1.8 eV and is 55\% at 3.1~eV.\cite{green2008self} Our calculations indicate a reflectance of  36\% at 1.8 eV, in excellent agreement with experiments. At 3.1~eV, however, our result of 64\% overestimates the experimental value.
The experimental reflectance of Si below the band gap of 1.12 eV is constant at 30\%.\cite{green2008self} This result compares well with our calculations for 400 $\mu$m thickness (Fig. \ref{fopt2}). Crystalline silicon presents an experimental transmittance of about 55\% between 0.5-1.0 eV\cite{serpenguzelsilicon} 
which agrees with our calculations shown in
Fig. \ref{fopt1}. These results leave a 15\% absorbance below the band gap which supports our way of calculating the solar cell efficiency
by taking the lower limit of the integral to be 0.31 eV (the beginning of AM1.5 data) instead of the band gap of the material as there is absorbance below the band gap. 
The highest experimental efficiency for Si is 24.7\%\cite{taguchi201324} for a 98 $\mu$m thick material while the Shockley-Quiesser limit value is 32.2\%.\cite{ruhle2016tabulated} We find an efficiency of 30.6\% for a 100 $\mu$m thick material for the ideal solar cell considered here.\cite{taguchi201324} 
There are a few considerations that explain the difference between the Shockley-Quiesser limit value and the efficiency value obtained in our work.
On one hand, in the Shockley-Quiesser limit the lower limit of the short-circuit current integral is taken to be the band gap of the material, however in our calculations, it is always 0.31 eV due to absorption below the band gap because the material is at room temperature which affects the tail of the imaginary part of the dielectric function below the optical gap\cite{lautenschlager1987temperature}.
On the other hand, in the Shockley-Quiesser limit 100\% light absorption is assumed above the band gap. However, we take reflections into account by introducing absorbance in the short-circuit integral thus lowering our calculated efficiency.

In the experiment,\cite{taguchi201324} the open-circuit voltage is measured to be 0.75 V which is significantly smaller than the value calculated here (0.922 V for the G$_0$W$_0$ band gap of 1.17 eV and 0.874 V for the experimental band gap of 1.12 eV,  for an ideal solar cell with a non-ideality factor of $n$=1). To obtain the experimental value for the open-circuit voltage, the non-ideality factor $n$ should be equal to 0.858 (here we used experimental band gap) for which we find a 25.6\% efficiency calculated with the experimental band gap of 1.12 eV and a 27.4\% efficiency calculated with the G$_0$W$_0$ band gap (Table \ref{tdata}). These calculations agree reasonably well 
with the experimental efficiency of 24.7\%. This value of efficiency is obtained for the dielectric function calculated with damping=0.01 (here damping is the Lorentzian broadening). This value of damping produces a tail of the dielectric function in better agreement with experiment (observe the behaviour at around 3 eV in Fig.\ref{feps}) as smaller damping values produce optical properties that are too oscillatory. 

Before moving to the nano scale results and the corresponding comparison with BP, it is worth noting the effect of the different approximations on the calculated efficiencies. 
If the lower limit of the short-circuit integral is taken to be the band gap instead of 0.31 eV, then the efficiency is found to be 17.4\% which is much lower than the experimental value. If the tail of the dielectric function were modeled better and better agreement of the dielectric function and the band gap with the experiment were obtained, then we would obtain efficiency values closer to the experiment. Furthermore, the experimental filling factor is 0.832 and the calculated value is 0.859. There is no significant difference in filling factors between theory and experiment.

\begin{figure*}[t]
\centering
\subfloat[]{\includegraphics[angle=0, scale=0.2]{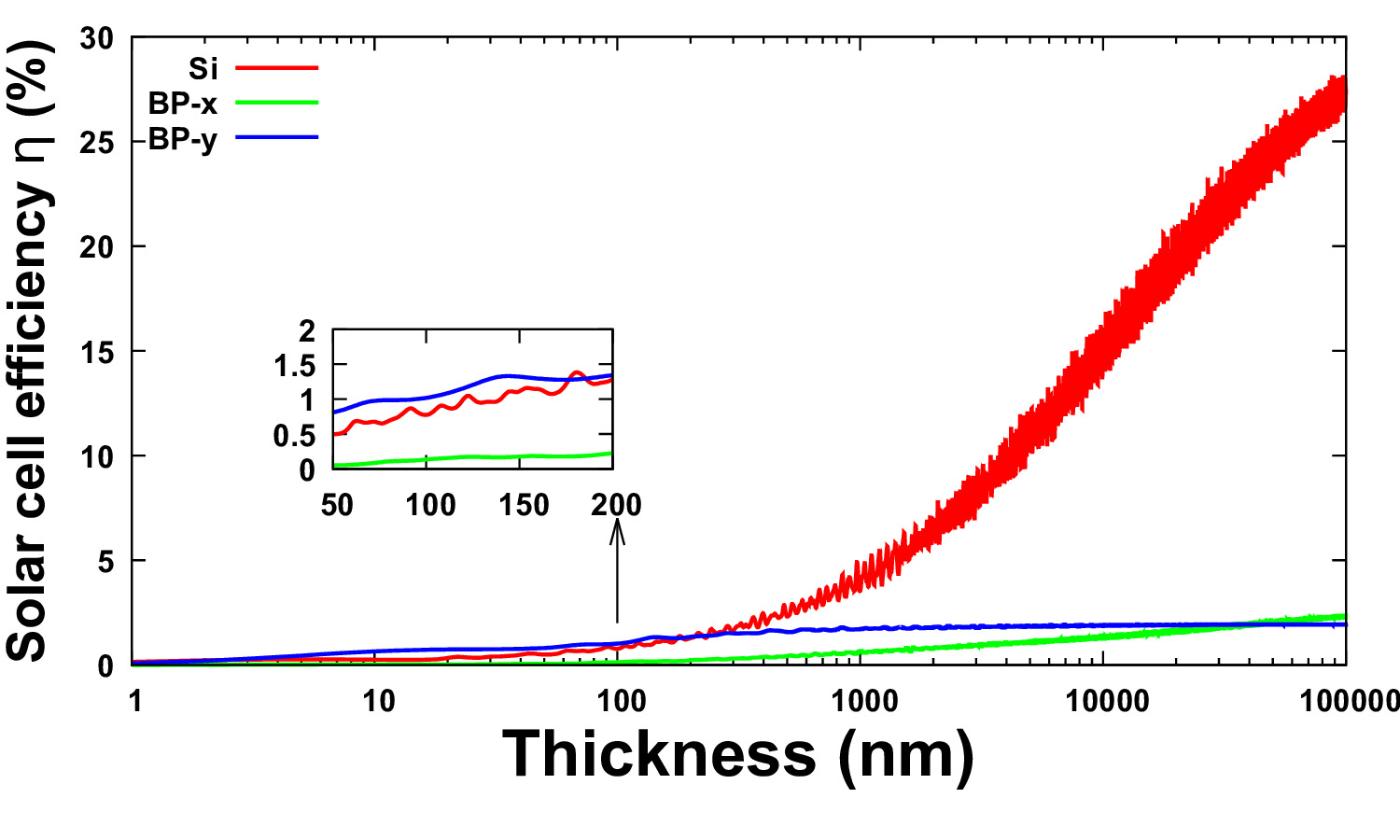}}
\caption{Comparison of solar cell efficiencies for Si and BP (for light polarized along the x and y directions) as a function of thickness from the nano to the micro scale.}
\label{fsolar}
\end{figure*}

We now turn to investigate the efficiency as a function of thickness. We find a 0.8\% efficiency for 100 nm thick Si (Table \ref{tdata}). This efficiency is much lower than the efficiency of transition metal dichalcogenides (about 20\%) at the same thickness as reported previously.\cite{ozdemir2020thickness} Moreover, at a 400 $\mu$m thickness the efficiency increases further to 30.1\%. However, in reality, Si presents defects that prevent excited electrons to diffuse long distances and recombine. Therefore, this efficiency cannot not achieved experimentally. The overall thickness-dependent solar cell efficiency of Si is shown in Fig. \ref{fsolar}.

\begin{figure*}[t]
\centering
\subfloat[]{\label{fepsBP6}\includegraphics[angle=0, scale=0.4]{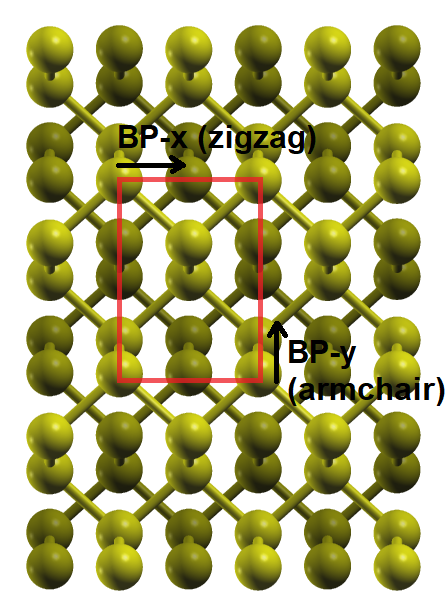}}
\subfloat[]{\label{fepsBP5}\includegraphics[angle=0, scale=0.15]{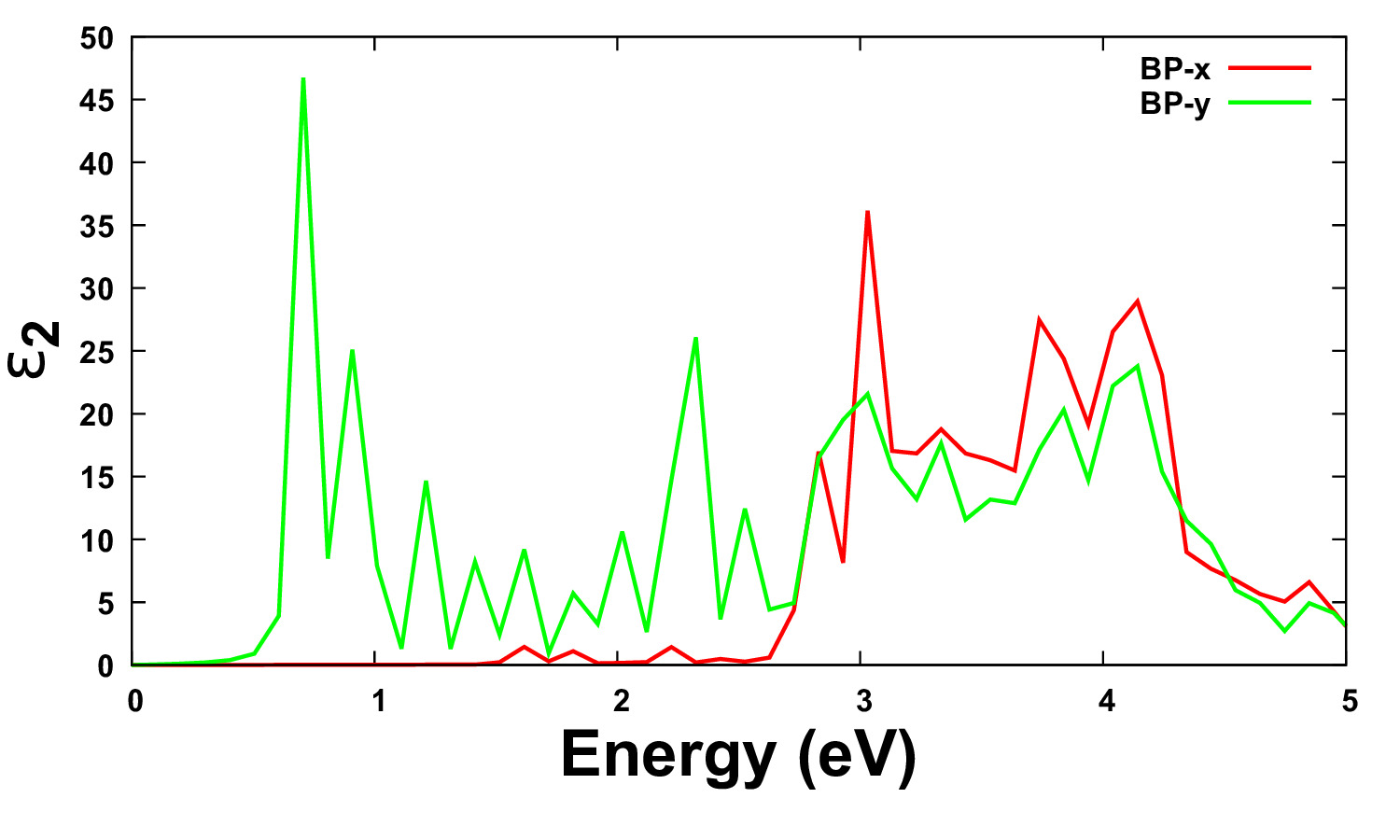}}
\caption{(a) Model of BP showing zigzag (BP-x) and armchair (BP-y) directions, (b) converged imaginary part of dielectric function ($\epsilon_2^{xx}$ and $\epsilon_2^{yy}$ elements) of BP from G$_0$W$_0$+BSE calculation for light polarized along the x and y directions where BP-x represents the zigzag direction and BP-y represents the armchair direction.}
\label{fepsBP}
\end{figure*}

\begin{figure*}[t]
\centering
\subfloat[]{\label{fepsBP1}\includegraphics[angle=0, scale=0.15]{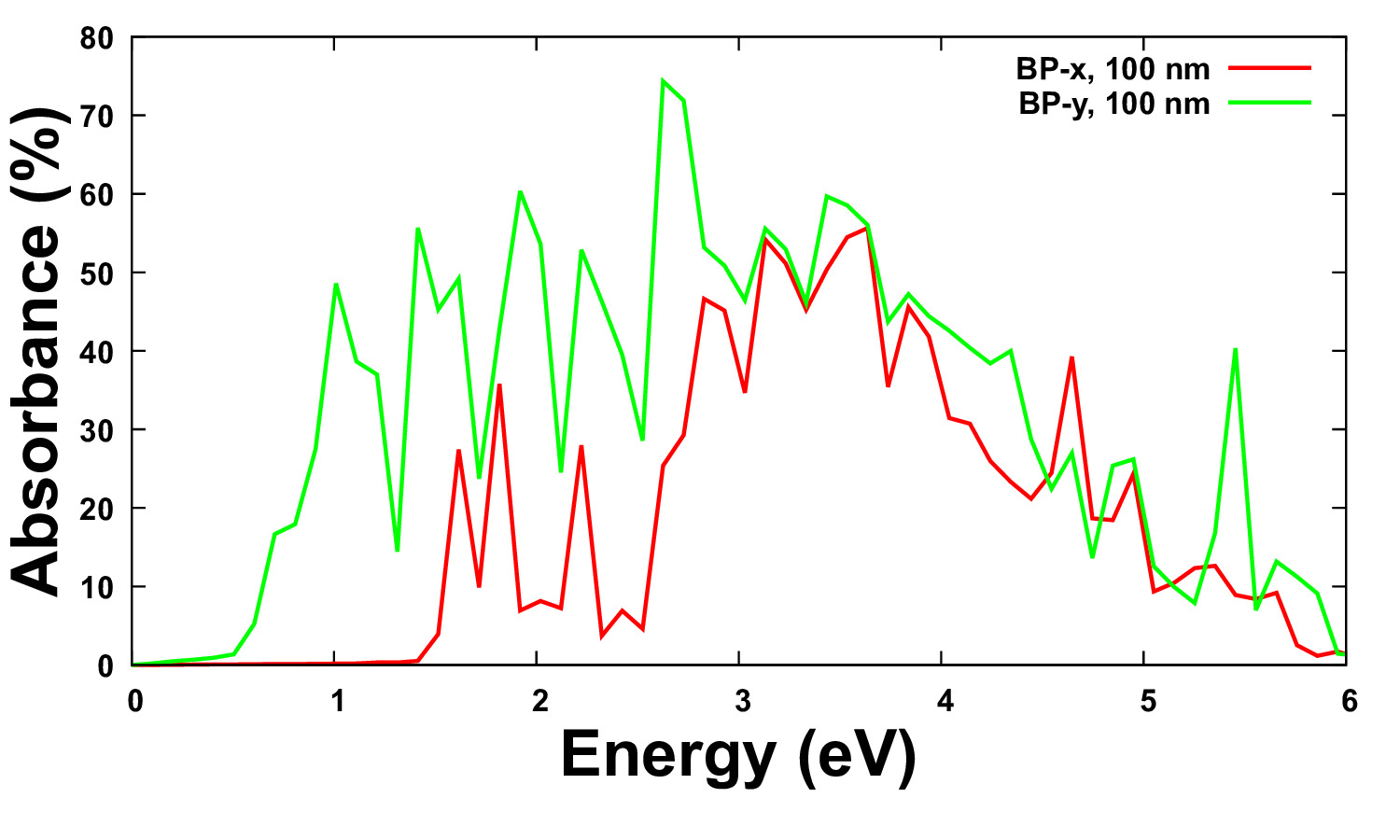}}
\subfloat[]{\label{fepsBP2}\includegraphics[angle=0, scale=0.15]{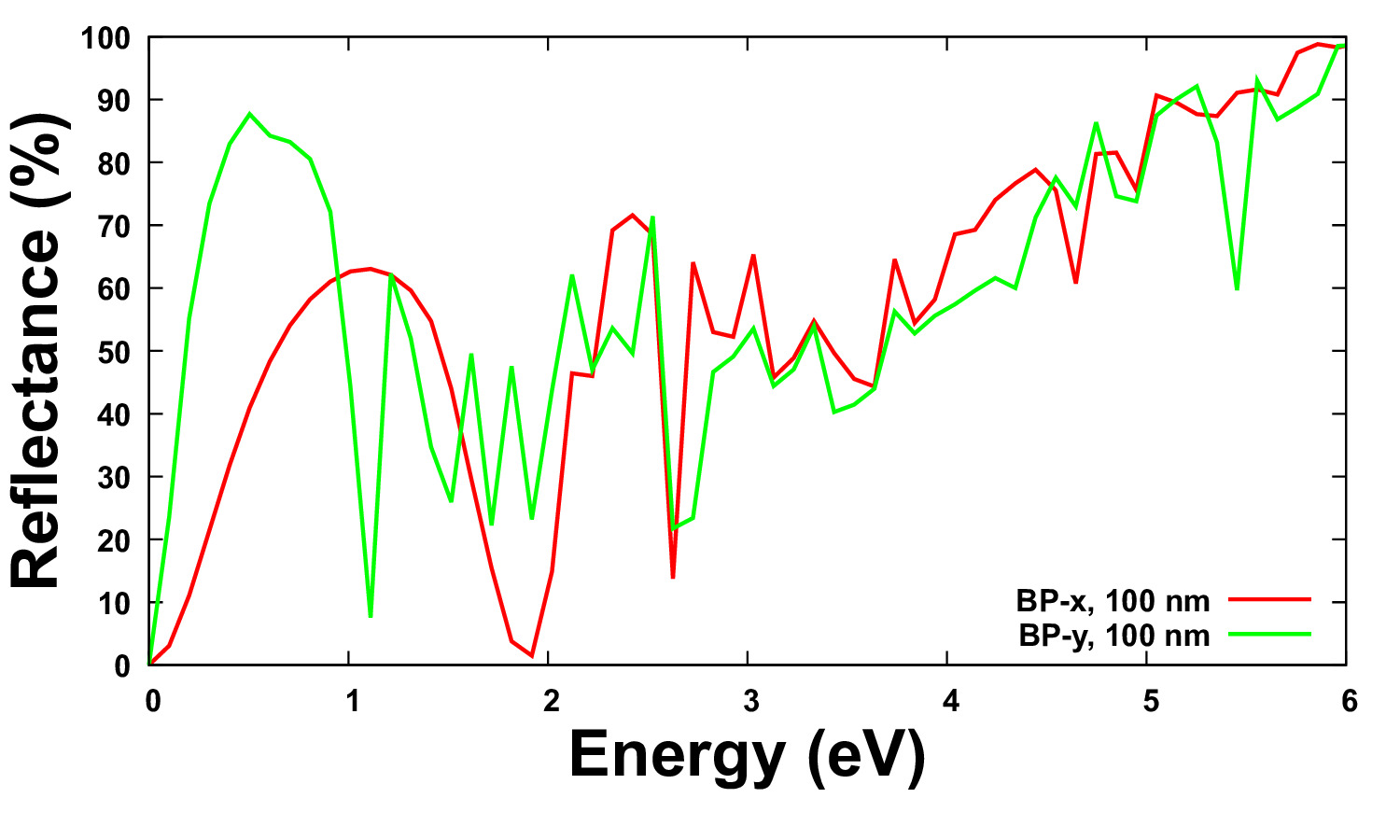}}\\
\subfloat[]{\label{fepsBP3}\includegraphics[angle=0, scale=0.15]{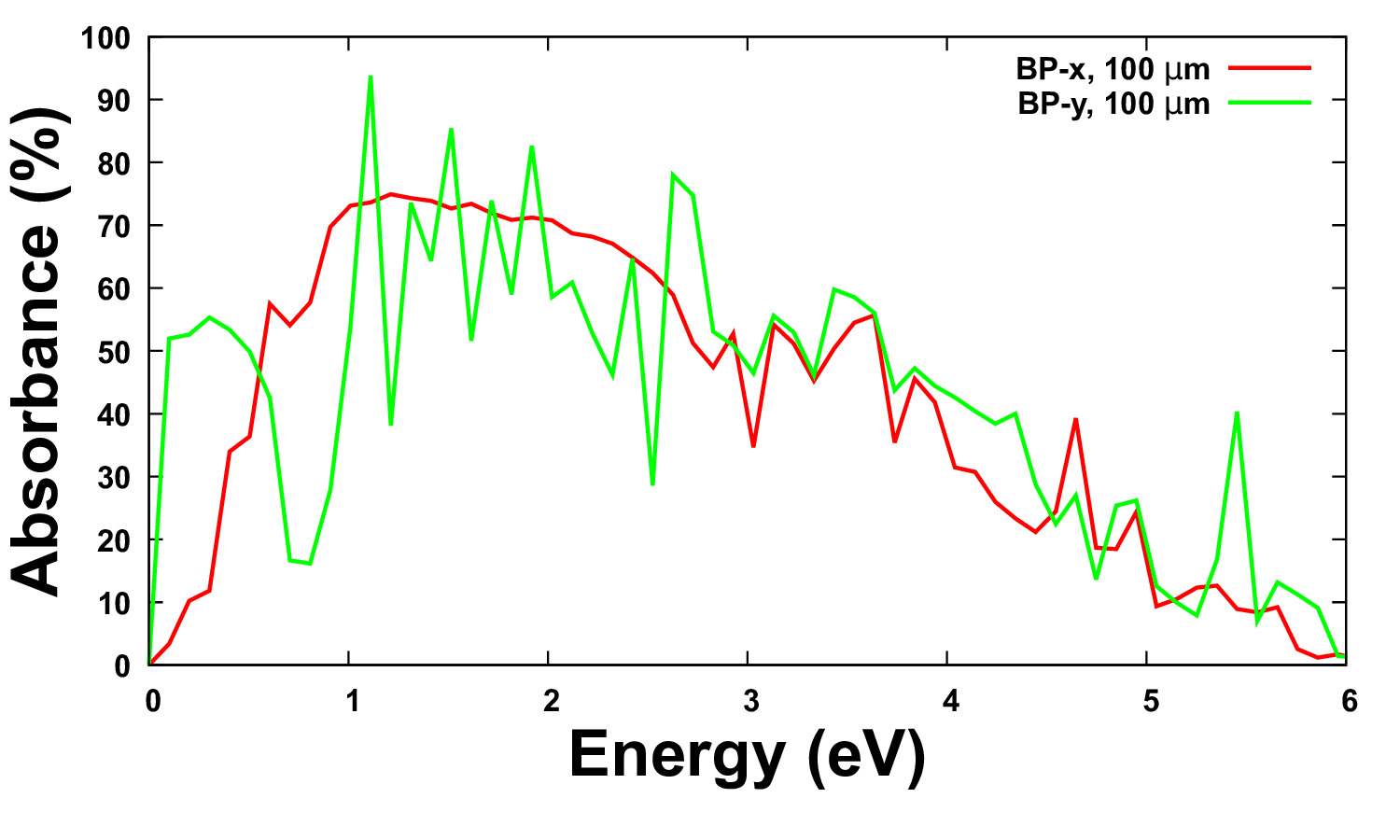}}
\subfloat[]{\label{fepsBP4}\includegraphics[angle=0, scale=0.15]{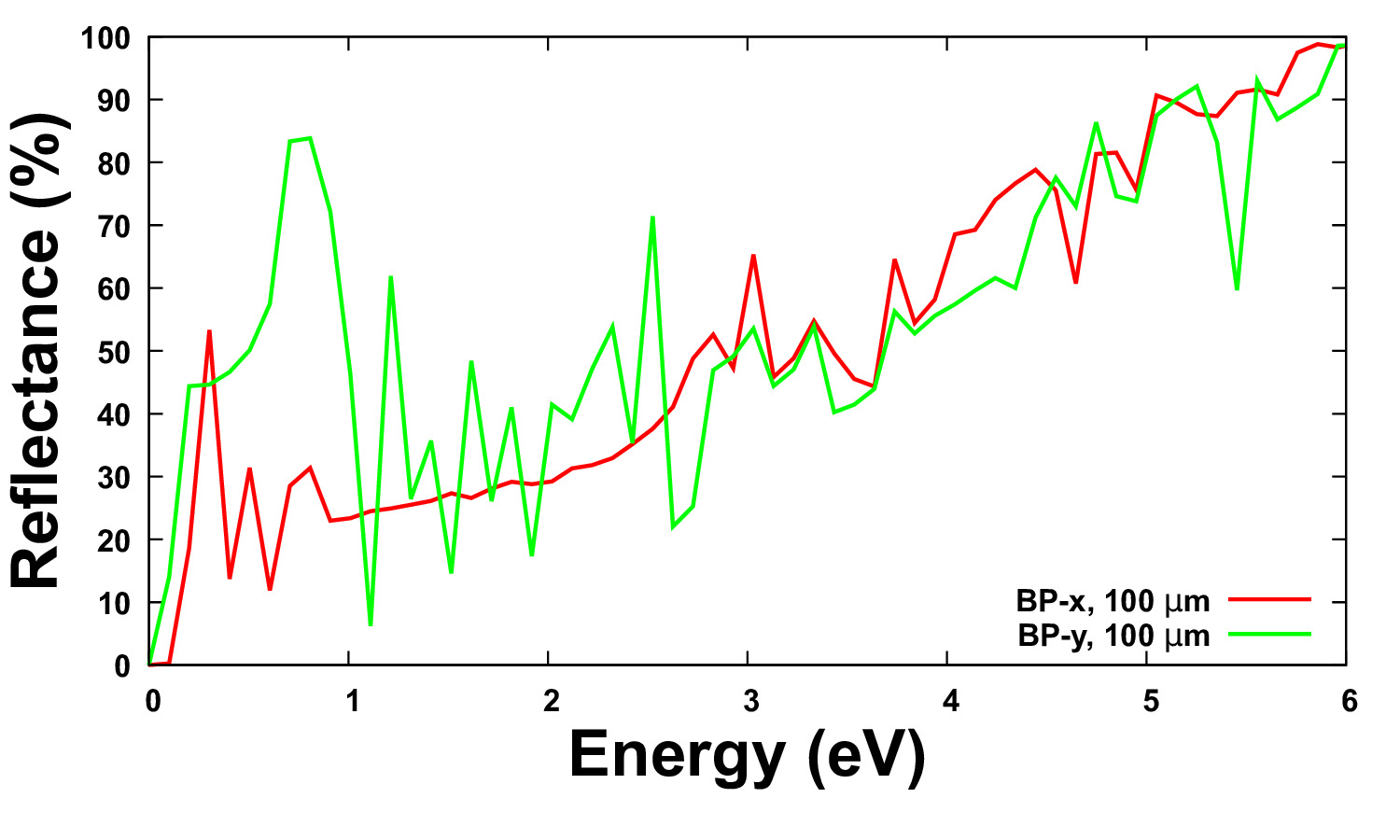}}\\
\caption{Absorbance for 100 nm thickness (a), reflectance for 100 nm thickness (b), absorbance for 100 $\mu$m thickness (c), and reflectance for 100 $\mu$m thickness (d) of BP for light polarized along the x and y directions.}
\label{fabsBP}
\end{figure*}

The same study performed for Si is carried out for BP. 
We use a 16$\times$12$\times$6 k-point grid for the density of states calculation as we find it to be a good compromise between accuracy and computational cost.
In this case, the total number of bands is converged at 200, the reduced block size is 450 and 4 valance bands and 5 conduction bands are used. The Fermionic temperature is set at 300 K and the damping is, again, chosen to be 0.01. Since BP is anisotropic, we calculated the solar cell efficiency for the light polarized along the x and y directions with BP-x representing the zigzag direction and BP-y representing the armchair direction (Fig. \ref{fepsBP6}). Absorbance and reflectance at two different thicknesses (100 nm and 100 $\mu$m) are shown in Fig. \ref{fabsBP}. From these results, it is clear that the optical properties of BP are significantly dependent on the polarization direction. Absorbance for 100 nm BP for light polarized along x-direction starts to increase at 1.5 eV, however, for light polarized along the y-direction starts to increase below 1 eV. These results imply that the solar cell efficiency for light polarized along the y-direction converges earlier than for light polarized along the x-direction. If we look at absorbance results for 100 $\mu$m, we observe that absorbance is slightly lower for light polarized along the y-direction indicating a lower solar cell efficiency. We calculate the solar cell efficiency of BP-x and BP-y to be 0.14\% and 1.02\%, respectively, using the experimental band gap of 0.3 eV for 100 nm thick BP. For this thickness, the solar cell efficiency of BP-y is larger than the corresponding efficiency of Si. If we look at thickness-dependent solar cell efficiency of BP-x and BP-y (see Fig. \ref{fsolar}), we observe that BP-y converges at a very low thickness in agreement with absorbance results. We obtain solar cell efficiencies of 2.33\% and 1.94\% for 100 $\mu$m thick BP along the x and y directions, respectively.

\begin{table}[t]
\caption{G$_0$W$_0$ band gap E$_g$, open-circuit voltage $V_{oc}$, filling factor FF, and solar cell efficiency $\eta$ for non-ideality factor $n=0.853$. For the calculation of the solar cell efficiency of BP, the experimental band gap of 0.3 eV is used.}
\begin{tabular}{c c c c c | c c c} \hline
&&\multicolumn{3}{c}{Thickness= 100 nm} &\multicolumn{3}{c}{Thickness = 100 $\mu$m}\\
\hline
&E$_g$&$V_{oc}$&FF&$\eta$ (\%)&$V_{oc}$&FF&$\eta$ (\%)\\
\hline\hline
Si (exp\cite{taguchi201324})&1.12&&&&0.750&0.832&24.7\\
Si&1.17&0.714&0.848&0.8&0.791&0.859&27.4\\
BP-x&0.74&0.060&0.402&0.14&0.105&0.502&2.33\\
BP-y&0.74&0.091&0.474&1.02&0.102&0.496&1.94\\
\hline
\end{tabular}
\label{tdata}
\end{table}

\section{Conclusion}

In summary, we have studied the optical properties and solar cell efficiencies of Si and BP using density functional theory. We find that it is paramount to correctly model the tail of the imaginary part of the dielectric function. Smaller damping values generally produce better results, however, if the damping is too small the optical properties become to be too oscillatory which returns inaccurate solar cell efficiencies. We find that a damping of 0.01 is suitable to calculate reasonable solar cell efficiencies. The solar cell efficiency of Si is calculated to be 27.4\% which is only slightly higher than the experimental value. Since BP has a small band gap, the solar cell efficiency of BP is calculated to be only 2.33\% for light polarized along the zigzag direction and 1.94\% for light polarized along the armchair direction for a 100 $\mu$m thickness. Notably, BP efficiency at low thicknesses for light polarized along the armchair direction is greater (1.02\% at 100 nm thickness) than the corresponding efficiency in Si. This results suggest that thin BP can be used as a part of a tandem solar cell to increase the overall efficiency.

\clearpage
\bibliographystyle{ieeetr}
\bibliography{main}
\end{document}